\headline={\ifnum\pageno>1 \hss \number\pageno\ \hss \else\hfill \fi}
\pageno=1
\nopagenumbers
\hbadness=100000
\vbadness=100000

\centerline{\bf FUNDAMENTAL WEIGHTS, PERMUTATION WEIGHTS}
\centerline{\bf AND WEYL CHARACTER FORMULA}
\vskip 15mm
\centerline{\bf H. R. Karadayi \footnote{*}{
e-mail: karadayi@itu.edu.tr} and M.Gungormez}
\centerline{Dept.Physics, Fac. Science, Tech.Univ.Istanbul }
\centerline{ 80626, Maslak, Istanbul, Turkey }
\vskip 10mm
\centerline{\bf{Abstract}}
\vskip 10mm

For a finite Lie algebra $G_N$ of rank N, the Weyl orbits $W(\Lambda^{++})$ of
strictly dominant weights $\Lambda^{++}$ contain $dimW(G_N)$ number of weights
where $dimW(G_N)$ is the dimension of its Weyl group $W(G_N)$. For any
$W(\Lambda^{++})$, there is a very peculiar subset $\wp(\Lambda^{++})$
for which we always have
$$ dim\wp(\Lambda^{++})=dimW(G_N)/dimW(A_{N-1}) \ \ . $$
For any dominant weight $ \Lambda^+ $, the elements of $\wp(\Lambda^+)$ are
called {\bf Permutation Weights}.

It is shown that there is a one-to-one correspondence between elements of
$\wp(\Lambda^{++})$ and $\wp(\rho)$ where $\rho$ is the Weyl vector of $G_N$.
The concept of signature factor which enters in Weyl character formula can
be relaxed in such a way that signatures are preserved under this one-to-one
correspondence in the sense that corresponding permutation weights have the
same signature. Once the permutation weights and their signatures are specified
for a dominant $\Lambda^+$, calculation of the character $ChR(\Lambda^+)$
for irreducible representation $R(\Lambda^+)$ will then be provided by $A_N$
multiplicity rules governing generalized Schur functions.
The main idea is again to express everything in terms of the
so-called {\bf Fundamental Weights} with which we obtain a quite relevant
specialization in applications of Weyl character formula. To provide
simplifications in practical calculations, a reduction formula governing
classical Schur functions is also given. As the most suitable one, the $E_6$
example which requires a sum over 51840 Weyl group elements, is studied
explicitly. This will be instructive also for an explicit application of
$A_5$ multiplicity rules.

In result, it will be seen that Weyl or Weyl-Kac character formulas find
explicit applications no matter how big is the rank of underlying algebra.

\vskip 15mm
\vskip 15mm
\vskip 15mm
\vskip 15mm
\vskip 15mm

\hfill\eject

\vskip 3mm
\noindent {\bf{I.\ INTRODUCTION}}
\vskip 3mm

It is well-known that summations over Weyl groups of Lie algebras enter in
many areas of physics as well as in mathematics. They are at the heart of all
character calculations for finite {\bf[1]} and also affine {\bf [2]} Lie
algebras and hence are of great importance in calculations of weight
multiplicities {\bf [3]} or in decompositions {\bf [4]} of tensor products
of irreducible representations. In high energy physics, it is known that
calculations of fusion coefficients {\bf[5]} or S-matrices which appear in
modular transformations {\bf[6]} of affine characters are directly related
to summations over Weyl groups. This however is not an easy task except for
a few cases which correspond to some Lie algebras of low rank.
Let us emphasize, for instance, that the summations are over respectively
51840, 2903040 and 696729600 Weyl group elements for $E_6, E_7$ and $E_8$
Lie algebras. It is therefore worthwhile to study the problem more closely.

In a previous work {\bf [7]} we have shown that in applications of Weyl
character formula for $A_N$ Lie algebras the sums over Weyl groups can be
represented by permutations. This in essence is being in line with the fact
that $A_N$ Weyl groups are already the permutation groups of (N+1) objects.
It is however interesting to note that this can be seen only when one uses
some properly chosen set of weights which we call {\bf fundamental weights}.
We have also shown that the signatures of Weyl reflections can then be
given precisely as being in relations with these permutations. One could
therefore expect that there is a way to extend this procedure to any other
finite Lie algebra ${\bf G_N}$ in view of the fact that it always has an
$A_{N-1}$ sub-algebra.

For this, let us recall that, one has, for any dominant weight $\Lambda^+$ of
${\bf G_N}$, a strictly dominant weight $\Lambda^{++} \equiv \Lambda^+ + \rho$
where $\rho$ is the Weyl vector of $G_N$. The character $ChR(\Lambda^+)$
of the corresponding irreducible representation $R(\Lambda^+)$ is then
given by
$$ ChR(\Lambda^+) = {A(\Lambda^{++}) \over A(\rho)}  \eqno(I.1) $$
\noindent where
$$ A(\mu) \equiv \sum_{\omega} \ \epsilon(\omega) \
e^{\omega(\mu)}  \eqno(I.2) $$
\noindent can be defined for any weight $\mu$. The sum here is over Weyl group
$W(G_N)$ and $\epsilon(\omega)$ is the signature of the Weyl reflection
$\omega$. The main emphasis now is on the fact that, for any strictly positive
dominant weight $\Lambda^{++}$, the number of elements of Weyl orbit
$W(\Lambda^{++})$ is always equal to the dimension of the corresponding Weyl
group. This hence allows us to re-write (I.2) in the form
$$ A(\Lambda^{++}) \equiv \sum_{\mu \in W(\Lambda^{++})} \ \epsilon(\mu) \
e^\mu  \eqno(I.3) $$
\noindent where $W(\Lambda^{++})$ is the corresponding Weyl orbit. One must
immediately note here that the concept of signature encountered in (I.2) is
conveniently relaxed in (I.3) in such a way that we introduce a signature
$\epsilon(\mu)$ for each and every weight $\mu$ within the Weyl orbit
$W(\Lambda^{++})$. It will be seen in the following that (I.3) is a quite
relevant form of (I.2) if one aims to apply it in Weyl character formula (I.1).
To this end, the concept of {\bf Permutation Weight} is of central importance.

\vskip 8mm
\noindent {\bf{II.\ PERMUTATION WEIGHTS}}
\vskip 3mm

It is known that the branching rules ${\bf G_N} \rightarrow A_{N-1}$ give us
irreducible $A_{N-1}$ representations which participate in the decomposition
of an irreducible representation of $G_N$. We, instead, want here to make
the same for Weyl orbits rather than representations.
For this, the following definition seems to be useful:
\par A Weyl orbit $W(\Lambda^+)$ always includes a sub-set $\wp(\Lambda^+)$
of weights having the form
$$ \sum_{i=1}^{N-1} \ k_i \ \lambda_i - k \ \lambda_N \ \ , \ \
k_i \in Z^+ \ \ , \ \ k \in Z \eqno(II.1) $$
\noindent where Z($Z^+$) is the set of integers(positive integers). The elements
of $\wp(\Lambda^+)$ are called {\bf permutation weights} of $\Lambda^+$.

\hfill\eject

$\lambda_I$'s and $\alpha_I$'s (I=1,2,..N) are respectively the fundamental
dominant weights and the simple roots of $G_N$. For details of Lie algebra
technology we refer to the excellent book of Humphreys {\bf [8]}.
As will be seen from the permutational lemma given in our previous
works {\bf [9]}, Weyl orbits of $A_N$ Lie algebras are stable under
permutations and this hence allows us to determine the complete weight
structure of an $A_N$ Weyl orbit. The permutation weights will give us
the same possibility but for any finite Lie algebra $G_N$ other than $A_N$
Lie algebras. We will therefore show now an explicit way to obtain all
permutation weights of a Weyl orbit $W(\Lambda^+)$ of $G_N$.

Let us first emphasize by definition that the sum of two permutation weights
is again a permutation weight. Let $\wp(\lambda)$ and $\wp(\lambda')$ be the
sets of permutation weights for $\lambda$ and $\lambda'$. It is then clear
that
$$ \wp(\lambda+\lambda') \subset \wp(\lambda) \cup \wp(\lambda') \eqno(II.2) $$
\noindent and for any element $\mu \in \wp(\lambda) \cup \wp(\lambda')$
one can also state $\mu \in \wp(\lambda+\lambda')$ on condition that
$$ (\mu,\mu) = (\lambda+\lambda',\lambda+\lambda') \eqno(II.3) $$
\noindent where (.,.) is the scalar product which can be introduced
on the weight lattice of ${\bf G_N}$. It is therefore sufficient to know
$\wp(\lambda_I)$'s (I=1,2,..N) in order to obtain the set $\wp(\Lambda^+)$ of
permutation weights for any dominant weight $\Lambda^+$ which is known to be
expressed by
$$ \Lambda^+ = \sum_{I=1}^N k_I \ \lambda_I \ \ , \ \ k_I \in Z^+ \ . $$

We find convenient here to exemplify our work in the Lie algebra of $E_6$
with the following Coxeter-Dynkin diagram:
$$ \hskip7.40cm 6 \hskip8cm $$
$  \hskip6cm 1 \hskip0.75cm 2 \hskip0.75cm 3 \hskip0.75cm 4 \hskip0.75cm 5 \hskip5cm  $

\vskip2mm

\noindent The permutation weight subsets of its fundamental Weyl orbits will
then be given by

$$ \eqalign{
\wp(\lambda_1) &\equiv \{ \lambda_1 \ , \ \lambda_1-\lambda_6 \ , \ \lambda_4-\lambda_6 \} \ , \cr
\wp(\lambda_2) &\equiv \{ \lambda_2 \ , \ \lambda_2 - 2 \lambda_6 \ , \
\lambda_3 + \lambda_5 - 2 \lambda_6 \ , \ \lambda_1 + \lambda_4 - \lambda_6 \ , \
\lambda_1 + \lambda_4 - 2 \lambda_6 \ , \ 2 \lambda_1 - \lambda_6 \} \ , \cr
\wp(\lambda_3) &\equiv \{ \lambda_3 \ , \
\lambda_1 + \lambda_3 + \lambda_5 - 3 \lambda_6 \ , \
\lambda_1 + \lambda_3 + \lambda_5 - 2 \lambda_6 \ , \
\lambda_2 + 2 \lambda_5 - 2 \lambda_6 \ , \cr
& \ \ \ \ \ \lambda_2 + \lambda_4 - 3 \lambda_6 \ , \
\lambda_2 + \lambda_4 - \lambda_6 \ , \ \lambda_3 - 3 \lambda_6 \ , \
2 \lambda_3 - 3 \lambda_6 \ , \ 2 \lambda_1 + \lambda_4 - 2 \lambda_6 \} \ , \cr
\wp(\lambda_4) &\equiv \{ \lambda_4 \ , \ \lambda_4 - 2 \lambda_6 \ , \
\lambda_1 + \lambda_3 - 2 \lambda_6 \ , \ \lambda_2 + \lambda_5 - \lambda_6 \ , \
\lambda_2 + \lambda_5 - 2 \lambda_6 \ , \ 2 \lambda_5 - \lambda_6 \} \ , \cr
\wp(\lambda_5) &\equiv \{ \lambda_5 \ , \ \lambda_2-\lambda_6 \ , \ \lambda_5-\lambda_6 \} \ , \cr
\wp(\lambda_6) &\equiv \{ \lambda_6 \ , \ -\lambda_6 \ , \
\lambda_1 + \lambda_5 - \lambda_6 \ , \ \lambda_3 - \lambda_6 \ , \ \lambda_3 - 2 \lambda_6 \} \ . }
\eqno(II.4)  $$

\noindent In the notation of $(k_1,k_2,k_3,k_4,k_5,k_6)$ for
$ \sum_{I=1}^6 k_I \ \lambda_I $, half of the 72 elements of $\wp(\rho)$ can
now be chosen, by direct use of (II.3), among elements of
$\sum_{I=1}^6 \ \wp(\lambda_I) $ as in the following:

$$ \eqalign{
\rho( 1) &= (1, 1, 1, 1, 1,  1) \ \ \ \ \ , \ \ \  \rho(13) = (3, 2, 2, 1, 2, -6) \ \ \ , \ \ \  \rho(25) = (3, 1, 3, 2, 1, -8)  \cr
\rho( 2) &= (1, 1, 2, 1, 1, -1) \ \ \ , \ \ \ \rho(14) = (2, 1, 2, 2, 3, -6) \ \ \ , \ \ \ \rho(26) = (6, 1, 1, 2, 1, -7)  \cr
\rho( 3) &= (1, 2, 1, 2, 1, -2) \ \ \ , \ \ \ \rho(15) = (1, 3, 1, 3, 1, -6) \ \ \ , \ \ \  \rho(27) = (1, 2, 1, 1, 6, -7)  \cr
\rho( 4) &= (1, 3, 1, 1, 2, -3) \ \ \ , \ \ \ \rho(16) = (4, 2, 1, 1, 3, -6) \ \ \ , \ \ \  \rho(28) = (2, 2, 2, 1, 4, -8)  \cr
\rho( 5) &= (2, 1, 1, 3, 1, -3) \ \ \ , \ \ \ \rho(17) = (3, 1, 1, 2, 4, -6) \ \ \ , \ \ \  \rho(29) = (4, 1, 2, 2, 2, -8)  \cr
\rho( 6) &= (2, 2, 1, 2, 2, -4) \ \ \ , \ \ \ \rho(18) = (4, 1, 3, 1, 1, -7) \ \ \ , \ \ \  \rho(30) = (2, 1, 4, 1, 2, -9)  \cr
\rho( 7) &= (1, 4, 1, 1, 1, -4) \ \ \ , \ \ \ \rho(19) = (1, 1, 3, 1, 4, -7) \ \ \ , \ \ \  \rho(31) = (7, 1, 1, 1, 1, -7)  \cr
\rho( 8) &= (1, 1, 1, 4, 1, -4) \ \ \ , \ \ \ \rho(20) = (2, 2, 2, 2, 2, -7) \ \ \ , \ \ \  \rho(32) = (1, 1, 1, 1, 7, -7)  \cr
\rho( 9) &= (3, 1, 2, 1, 3, -5) \ \ \ , \ \ \ \rho(21) = (5, 1, 2, 1, 2, -7) \ \ \ , \ \ \  \rho(33) = (1, 3, 1, 1, 5, -8)  \cr
\rho(10) &= (2, 3, 1, 2, 1, -5) \ \ \ , \ \ \ \rho(22) = (2, 1, 2, 1, 5, -7) \ \ \ , \ \ \  \rho(34) = (5, 1, 1, 3, 1, -8)  \cr
\rho(11) &= (1, 2, 1, 3, 2, -5) \ \ \ , \ \ \ \rho(23) = (3, 2, 1, 2, 3, -7) \ \ \ , \ \ \  \rho(35) = (3, 1, 3, 1, 3, -9)  \cr
\rho(12) &= (4, 1, 1, 1, 4, -5) \ \ \ , \ \ \ \rho(24) = (1, 2, 3, 1, 3, -8) \ \ \ , \ \ \  \rho(36) = (1, 1, 5, 1, 1, -10)    } \eqno(II.5) $$

\hfill\eject

\vskip 8mm
\noindent {\bf{III.\ EXPLICIT CONSTRUCTION OF WEYL ORBITS}}
\vskip 3mm

It is known that the complete set of weights of a Weyl orbit is obtained by
the fact that Weyl orbits are by definition stable under Weyl reflections.
Instead, we want to construct Weyl orbits here by knowing their permutation
weights solely. As in above, let $\lambda_I$'s be the fundamental dominant weights
of ${\bf G_N}$ whereas $\sigma_i$'s be the ones for its $A_{N-1}$ sub-algebra.

The existence of such a sub-algebra can always be shown explicitly by taking
$$ \sigma_i = \lambda_i - n_i \lambda_N  \eqno(III.1) $$
\noindent where $n_i$'s are some specified rational numbers. Let us recall
from our previous references {\bf [7, 9]} that fundamental weights
$\mu_I$ (I=1,2,..N) for $A_{N-1}$ sub-algebra are defined by
$$ \sigma_i \equiv \mu_1 + \mu_2 + .. + \mu_i  \ \ , \ \  i=1,2,..,N-1.  \eqno(III.2) $$
\noindent together with the condition that
$$ \mu_1 + \mu_2 + .. + \mu_N \equiv 0 \eqno(III.3)  $$
\noindent and also
$$ (\mu_I,\lambda_N) \equiv 0 \ \ .  \eqno(III.4) $$
\noindent The permutational lemma then states for an $A_{N-1}$ dominant weight
$$ \sigma^+ = s_1 \mu_1 + s_2 \mu_2 + .. + s_N \mu_N \ \ , \ \
s_1 \geq s_2 \geq .. \geq s_N \geq 0   \eqno(III.5)  $$
\noindent that its Weyl orbit $W(\sigma^+)$ are obtained to be
$$ W(\sigma^+) = \{ s_1 \mu_{I_1} + s_2 \mu_{I_2} + .. + s_N \mu_{I_N} \}
 \eqno(III.6) $$
\noindent by permutating fundamental weights $\mu_I$'s.
Note here that no two of indices $I_1,I_2,..,I_N$ (=1,2,..,N) shall take the
same value. This is also true for all permutation weights because,
for $\lambda_N \rightarrow 0$, they turn out to be $A_{N-1}$ dominant weights.
We then obtain an extension of the permutational lemma for any finite
Lie algebra other than $A_N$ Lie algebras.

An example will again be helpful here. Let us consider $E_6 \rightarrow A_5$
decomposition which is specified by
$$
\sigma_1 = \lambda_1 -{1 \over 2} \lambda_6 \ , \
\sigma_2 = \lambda_2 -{2 \over 2} \lambda_6 \ , \
\sigma_3 = \lambda_3 -{3 \over 2} \lambda_6 \ , \
\sigma_4 = \lambda_4 -{2 \over 2} \lambda_6 \ , \
\sigma_5 = \lambda_5 -{1 \over 2} \lambda_6  \eqno(III.7) $$

\noindent where $\lambda_I$'s (I=1,2,..6) are $E_6$ fundamental dominant
weights while $\sigma_i$'s (i=1,2,..5) are those of $A_5$.
\noindent The influence of $A_5$ permutational lemma for $E_6$ Weyl orbits
can be illustrated, in view of (II.4), in the following example:
$$ W(\lambda_1) = \{ W(\sigma_1) +  {1 \over 2} \Omega \ , \
W(\sigma_1) - {1 \over 2} \Omega \ , \
W(\sigma_4) \} \eqno(III.9) $$
\noindent where, for $A_5$ Weyl orbits, we know that
$$ W(\sigma_1) = \{ \mu_{I_1} \} \ , \
W(\sigma_4) = \{ \mu_{I_1} + \mu_{I_2} + \mu_{I_3} + \mu_{I_4} \} \ , \
I_1 \geq I_2 \geq I_3 \geq I_4 = 1,2,..6 \ .  $$
\noindent In (III.9), one keeps the notation $\Omega \equiv \lambda_6$ for
which we know  that $(\Omega , \mu_I) = 0$. It is, in fact, nothing but an
example of the {\bf branching rule of Weyl orbits} which is at the heart of
our definition of permutation weights. The branching rules for the remaining
fundamental $E_6$ Weyl orbits $W(\lambda_i)$ for i=2,3,..,6 can be obtained
similarly from the permutation weights given in (II.4).

What we want to emphasize here is mainly that 72 permutation weights of
$W(\rho)$ of $E_6$ will be given by
$$ \wp(\rho) = ( \sigma(k)^{++} + r(k) \ \Omega \ , \
\sigma(k)^{++} - r(k) \ \Omega )  \eqno(III.10)  $$
\noindent where the $\Omega$ extension parameters r(k)'s are some
{\bf positive} rational numbers. 36 strictly dominant weights $\sigma(k)^{++}$
and their parameters r(k)'s can be determined from (II.5) respectively
(k=1,2,...36).

\vskip 8mm
\noindent {\bf{IV.\ APPLICATIONS OF WEYL FORMULA AND A LEMMA }}
\vskip 3mm

Since we have in mind to make summations over Weyl groups explicitly, the
decomposition (III.10) gives us for $E_6$ the possibility to calculate
$A(\rho)$ by the aid of (I.3). Let us recall from ref.{\bf [7]} that for
anyone of the $A_5$ dominant weights
$$\sigma^{++}(k) \equiv \sigma^+(k) + \rho_\sigma$$
\noindent participated in the list (III.10), one has
$$ A(\sigma^{++}(k) + r(k) \ \Omega) = A(\rho_\sigma) \ S(\sigma^+(k))
\ u^{r(k)} \eqno(IV.1) $$
\noindent where $\rho_\sigma$ is the Weyl vector of $A_5$. In the specialization
$$ e^\Omega \equiv u \ , \ e^{\mu_I} \equiv u_I \ , \ I=1,2,..6 \eqno(IV.2) $$
\noindent of formal exponentials, we know that $S(\sigma^+(k))$ is a
generalized Schur function {\bf [7, 10]} which can be reduced via $A_5$
multiplicity rules to a polynomial expression in terms of 5 indeterminates
$x_i$ (i=1,2,..5) which are defined by
$$ u_1^M + u_2^M + u_3^M + u_4^M + u_5^M + u_6^M \equiv M \ x_M \ \ ,
\ \  M=1,2,...  \ \ \eqno(IV.3)  $$
\noindent Note here as a result of (III.3) that 6 indeterminates $u_I$ are
constrained by
$$ \prod_{I=1}^6 \ u_I \equiv 1  \eqno(IV.4) $$
and hence one can immediately see from definitions (IV.3) that, for $M>5$,
all indeterminates $x_M$ depend non-linearly on the first five indeterminates
$x_i$ (i=1,2,..5). This will also give rise to some reduction
rules governing classical Schur functions {\bf [7]} which are defined by
$$ S(M \ \lambda_1) \equiv S_M(x_1,x_2,..,x_5) \ \ , \ \ M=1,2,..5,6,..
\eqno(IV.5) $$
\noindent where $S_M(x_1,x_2,..,x_5)$'s are some polynomials which can be
obtained for M=1,2,..5 directly. For \break $M>5$, however, one must take into
account the above mentioned non-linear relations among indeterminates $x_M$.
Practical calculations could get complicated in general for $A_N$ multiplicity
rules. For this, we find convenient to give here some clarifying details.
It will be seen in fact that these non-linear relations governing indeterminates
$x_M$ for $M>N$, result in the following reduction rules among polynomials
$ S_M(x_1,x_2,..,x_N) \equiv S_M(N) $ which correspond to classical Schur
functions as in (IV.5):
$$ S_M(N) = (-1)^N \ S_{M-N-1}(N) - \sum_{i=1}^N S_i^{*}(N) \  S_{M-i}(N) \ \ ,
\ \  M>N \eqno(IV.6) $$
\noindent where $S_M^{*}(N)$ is obtained from $S_M(N)$ under the replacements
$x_i \rightarrow - \ x_i$. It will be seen that the reduction rules given in
(IV.6) prove extremely useful in applications of $A_N$ multiplicity rules
especially for higher values of the rank N.

Another important notice here is to give a precise definition of signatures for
72 permutation weights participated in the decomposition (III.10). The
arrangement in (III.10) is in such a way that
$$ \eqalign{
\epsilon(\sigma^{++}(k) + r(k) \ \Omega) &\equiv +1  \cr
\epsilon(\sigma^{++}(k) - r(k) \ \Omega) &\equiv -1   } \eqno(IV.7) $$
\noindent for  k=1,2,..36. The miraculous factorization (IV.1) of the Weyl
formula comes out only by the aid of such a choice.

It is thus seen that the decomposition (III.10) of $\wp(\rho)$ allows us
to calculate $A(\rho)$ but nothing says about any other $A(\Lambda^{++})$
which we need in the calculation of the character $ChR(\Lambda^+)$. For this,
a lemma which assures one-to-one correspondence between 72 elements of
(III.10) and those of any other $\wp(\Lambda^{++})$ would be of great help.
In view of the condition (II.3), there is a one-to-one correspondence
which maps any element of $\wp(\rho)$ to one and only one element of
$\wp(\Lambda^{++})$ in such a way that their signatures are preserved.
The generalization leads us to the following {\bf lemma}:

Let, for any dominant weight $\Lambda^+$, $\wp(\Lambda^+)$ be the
subset of its permutation weights and also
$$ \wp(\rho) \equiv \{ \rho(k) \} \ \ , \ \
\wp(\Lambda^{++}) \equiv \{ \Lambda(k) \}  \eqno(IV.8) $$
\noindent for any Lie algebra $G_N$ with the Weyl vector $\rho$. Then,
in view of condition (II.3), for
$$ k=1,2,... dimW(G_N)/dimW(A_{N-1}) $$
\noindent and $ \mu \in \wp(\Lambda^+) $ there is a one-to-one correspondence
$\Xi$ which provides
$$ \Xi \ : \ \rho(k) + \mu \rightarrow \Lambda(k) \eqno(IV.9) $$
\noindent in such a way that
$$ \epsilon(\Xi(\rho(k))) \equiv \epsilon(\rho(k)) \ \ . \eqno(IV.10) $$
Note here that, we always have
$$ dim\wp(\Lambda^{++}) \geq dim\wp(\Lambda^+)  $$
\noindent and for each and every value of k there is one and only one
$ \mu \in \wp(\Lambda^+) $.

In the conclusion, we can say that the decomposition (II.5) makes any explicit
summation over 51840 elements of $E_6$ Weyl group possible and hence completely
solves the problem for $E_6$ Lie algebra. One must however add that related
definitions must be made precisely case by case for any other Lie algebra.
For all the chains $B_N , C_N , D_N$, the exceptional Lie algebras $G_2 , F_4$
and even for $E_7$ the method presented above is easily tractable as we will
show in a subsequent paper. The same could also be true for $E_8$ but again
one must note that we have $ dim\wp(\rho) = 17280 $ for $E_8$. We finally
remark that a similar analysis can be presented in the framework of $A_8$
sub-algebra of $E_8$ and this makes the problem more tractable by reducing
the number of permutation weights down to 1920. To the knowledge of authors,
this is quite convenient to handle any problem which requires summations over
696729600 elements of $E_8$ Weyl group in an explicit manner and hence it
would be worthwhile to study in another publication.

As the last but not the least, let us add that all these calculations can be
performed by the aid of very simple computer programs, say, in the language
of Mathematica {\bf [11]}.

\vskip3mm
\noindent{\bf {REFERENCES}}
\vskip3mm

\leftline{[1] H.Weyl, The Classical Groups, N.J. Princeton Univ. Press (1046)}

\leftline{[2] V.G.Kac, Infinite Dimensional Lie Algebras, N.Y., Cambridge Univ. Press (1990)}

\leftline{[3] H.Freudenthal, Indag. Math. 16 (1964) 369-376}
\leftline{\ \ \ \ G.Racah, Group Theoretical Concepts and Methods in Elementary Particle Physics,}
\leftline{\ \  \ \ ed. F.Gursey, N.Y., Gordon and Breach (1964) }
\leftline{\ \ \ \  B.Kostant, Trans.Am.Math.Soc., 93 (1959) 53-73}

\leftline{[4] A.U.Klimyk, Amer.Math.Soc.Translations, Series 2,vol.76, (1968) Providence}
\leftline{\ \ \ \ R.Steinberg, Bull.Amer.Math.Soc., 67 (1961) 406-407}

\leftline{[5] E.Verlinde, Nucl.Phys. B300 (1988) 360 }
\leftline{\ \ \ \ J.Fuchs, Fortsch.Phys. 42 (1994) 1-48}

\leftline{[6] V.G.Kac and S.Peterson, Adv. Math., 53 (1984) 125-264}

\leftline{[7] H.R.Karadayi, $A_N$ Multiplicity Rules and Schur Functions, submitted for publication}

\leftline{[8] J.E.Humphreys, Introduction to Lie Algebras and Representation Theory, N.Y., Springer-Verlag (1972)}

\leftline{[9] H.R.Karadayi and M.Gungormez, Jour.Math.Phys., 38 (1997) 5991-6007}
\leftline{ \ \ \ \ H.R.Karadayi, Anatomy of Grand Unifying Groups I and II ,}
\leftline{ \ \ \ \ ICTP preprints(unpublished) IC/81/213 and 224}

\leftline{[10] I.G.MacDonald, Symmetric Functions and Hall Polynomials,
Clarendon Press (1995), Oxford }

\leftline{[11] S. Wolfram, Mathematica$^{TM}$, Addison-Wesley (1990) }

\end